\documentclass[preprint]{aastex6}
\usepackage{xcolor}
\usepackage{graphicx,color}
\usepackage{amsmath,amssymb}
\usepackage{hyperref}
\usepackage{epstopdf}
\usepackage{ulem}
\usepackage[T1]{fontenc}
\usepackage{newunicodechar}
\usepackage[utf8]{inputenc}

\usepackage{rotating}

\newcommand{\gev}{\ensuremath{\,\mathrm{GeV}}}
\newcommand{\tev}{\ensuremath{\,\mathrm{TeV}}}
\newcommand{\kpc}{\ensuremath{\,\mathrm{kpc}}}
\newcommand{\ev}{\ensuremath{\,\mathrm{eV}}}
\newcommand{\gyr}{\ensuremath{\,\mathrm{Gyr}}}

\newcommand{\gau}{\ensuremath{\mathcal{G}(\boldsymbol{r}-\boldsymbol{r_i})}}

\begin{document}



\title{Ultra-Light Axion Dark Matter and 
its impacts on dark halo structure in $N$-body simulation}

\author{Jiajun Zhang}
\affil{School of Physics and Astronomy, Shanghai Jiao Tong University, Shanghai, China}
\affil{Department of Physics, The Chinese University of Hong Kong, Hong Kong, China}
\author{Yue-Lin Sming Tsai}
\affil{Physics Division, National Center for Theoretical Sciences, 
Hsinchu, Taiwan}
\author{Jui-Lin Kuo}
\affil{Department of Physics, National Tsing Hua University, Hsinchu, Taiwan}
\author{Kingman Cheung}
\affil{Physics Division, National Center for Theoretical Sciences, 
Hsinchu, Taiwan}
\affil{Department of Physics, National Tsing Hua University, Hsinchu, Taiwan}
\affil{Division of Quantum Phases and Devices, School of Physics, 
Konkuk University, Seoul 143-701, Republic of Korea}
\author{Ming-Chung Chu}
\affil{Department of Physics, The Chinese University of Hong Kong, Hong Kong, China}




\begin{abstract}
The Ultra-Light Axion (ULA) is a dark matter candidate with mass
$\mathcal{O}(10^{-22})\ev$ and de Broglie wavelength of order
$\kpc$. Such an axion, also called the Fuzzy Dark Matter (FDM), thermalizes via
the gravitational force and forms a Bose-Einstein condensate.  
Recent studies suggested that the quantum pressure from the FDM
can significantly affect the structure formation in small scales, thus
alleviating the so-called ``small-scale crisis''.
In this paper, we develop a new technique to discretize the 
quantum pressure and illustrate the interactions among FDM particles 
in the $N$-body simulation, which accurately simulates the
formation of the dark-matter halo and its inner structure 
in the region outside the softening length. 
In a self-gravitationally-bound virialized halo, we find a constant density,
solitonic core, which is consistent with the theoretical prediction. 
The existence of the solitonic core reveals the non-linear effect of quantum pressure 
and impacts the structure formation in the FDM model.
\end{abstract}




%
\section{Introduction}
There have been many compelling evidences for the existence of 
cold dark matter (CDM), which successfully explains the rotation 
curves of spiral galaxies, the Cosmic Microwave Background power spectrum, the Bullet Cluster, and 
large scale structure formation of the Universe.
%
The most well-motivated model for CDM is the weakly interacting massive 
particles (WIMPs), whose mass range is from the sub-GeV to $100\tev$ 
owing to the relic density requirement. 
Unfortunately, up to now no compelling evidence of WIMPs was found in 
all different types of searches, such as collider 
searches~\citep{Aaboud:2016tnv,CMS:2016tns}, 
underground detections~\citep{Akerib:2016vxi,tan2016dark}, and 
astronomical observations~\citep{ackermann2015searching,Aartsen:2016pfc}. 
Null signals reported in all these experiments have 
shrunk the parameter space of many WIMP models to some finely-tuned regions. 
Also, the next generation of CDM searches are shifting their focuses 
to different mass regions. 

The CDM model through detailed $N$-body simulations, 
though successfully explains the observations
in large scales, fails to account for the observations in 
relatively smaller scales; this is known as the ``small-scale crisis'':
(i) the missing satellites problem~\citep{Moore:1999nt,Klypin:1999uc},
(ii) the cusp-core problem~\citep{deBlok:2009sp}, and
(iii) the too-big-to-fail problem~\citep{BoylanKolchin:2011de}.
In order to alleviate the problems, two types of mechanism are introduced.
Some authors believe that the problems can be alleviated by considering baryonic feedback
carefully~\citep{Schaller:2014uwa,Schaller:2014gwa}.
Others believe that a new mechanism of velocity boost 
is needed for the DM momentum exchange beyond the collisionless picture of
the CDM. Examples include 
strongly self-interacting DM~\citep{Tulin:2013teo}, Fuzzy Dark Matter (FDM)
\citep{Hu:2000ke}, etc. Both of them share the the feature of smoothing out the cuspy 
matter distribution. 

The Ultra-Light Axion (ULA) or so-called Fuzzy Dark Matter (FDM) 
is then a good candidate for CDM~\citep{Turner:1983he,Press:1989id,Sin:1992bg,Goodman:2000tg,Peebles:2000yy,matos2002scalar,guzman2003newtonian,amendola2006dark,calabrese2016ultra,kim2016ultralight}. 
It not only keeps the 
success of CDM in dealing with large-scale issues, 
but also provides a possible solution to the small-scale crisis.
The FDM is a scalar boson with an extremely light mass $\gtrsim 10^{-22}\ev$,
which is required by a recent observation on
the reionization history of the Universe~\citep{bozek2015galaxy}. 
A recent discussion about the constraints on the FDM mass from CMB can be found in
Ref.~\citep{Hlozek:2016lzm}.
With a common velocity in the order of 100 km s$^{-1}$,   
the de Broglie wavelength of the FDM is very long $\sim\mathcal{O}(\kpc)$ 
~\citep{Sin:1992bg,Hu:2000ke,Ferrer:2004xj,Boehmer:2007um,
Lee:2008jp,Sikivie:2009qn,Mielke:2009zza,chavanis2011analytical,chavanis2011numerical,dev2016gravitational,
marsh2016axion}. 
Some authors suggested that the FDM is in Bose-Einstein Condensate (BEC)~\citep{schive2014cosmic}. 
However it is still an open question, because the BEC transition temperature 
is higher than the cosmological temperature. 
Hence, in this paper we will not join such a debate but instead focus on numerically solving 
the structure formation of the FDM model efficiently.
The FDM is non-thermally produced and thus 
still in the non-relativistic regime and behaves like CDM. 
However, due to the quantum nature of the FDM in small scales, about tens or hundreds of $\kpc$, 
remarkably the FDM is distinguishable from the normal CDM.
One can find those differences in 
the matter power spectrum~\citep{veltmaat2016cosmological}, 
halo mass function (HMF)~\citep{du2016substructure}, and 
halo structure~\citep{schwabe2016simulations}. 
The most striking feature of the FDM is that the halo has a solitonic 
core of size $\sim\mathcal{O}(\kpc)$ resulting from the 
quantum pressure of the FDM particles, which can be larger than
their self-gravity. 
Hence, the quantum pressure plays an essential role in solving 
the small-scale crisis. Moreover, if the halo small scale structure 
can be measured more accurately in the future, 
the FDM particle mass can be constrained.
The latest reviews are given in Ref.~\citep{marsh2016axion,Hui:2016ltb}.

In this work, we propose a new scheme -- the effective Particle-Particle (PP) 
interaction -- for simulating the FDM model, by which one can compute the quantum 
effect of the FDM in the $N$-body simulation with high resolution. 
In a self-gravitationally-bound virialized halo, we find 
a constant density core -- solitonic core -- of size of around 
$1\kpc$, but with a lower density than that in the conventional 
CDM model.
The result shows the non-trivial quantum pressure effect on the structure formation.
We present the effects of the linear and non-linear power spectrum growth, 
especially for the non-linear effect in high density regions. 
For small scale structures, particularly for scales smaller (larger) than one de Broglie wavelength,  
the quantum pressure is attractive (repulsive). This leads to non-trivial structure formation in 
high density regions. 
However, in early age, low over-density regions, 
the linear effect is mainly from the repulsive quantum pressure 
which suppresses the matter power spectrum at scales smaller than the Jeans scale.  

We have developed an independent and completely different
simulation scheme compared to a few previous studies, 
which were based on direct cosmological simulations with the Schr{\"o}dinger-Poisson 
equations~\citep{schive2014cosmic}, Smoothed Particle Hydrodynamic
(SPH) scheme simulation technique~\citep{mocz2015numerical} and Particle-Mesh (PM) scheme 
simulation technique~\citep{veltmaat2016cosmological}. 
We cross-check our results with previous simulations and find them consistent with each other. 
Nevertheless, our implementation of the quantum effect with a simple PP method helps us 
to explain some of the previously unclear behavior and understand self-consistently 
how quantum pressure affects the structure formation.

The main advantage of our scheme is its high running efficiency. 
From all the tests that we have performed, the CPU time required for the FDM 
simulations is only two to three times longer than the corresponding CDM simulations,
which is well expected for our method described here. Thus, we can perform cosmological
simulation in larger scale than previous literature with the same amount of computational resources
~\citep{schive2014cosmic,mocz2015numerical,veltmaat2016cosmological}.

The following sections are arranged as follows. 
In Sec.~\ref{sec:method}, we introduce the theoretical approach 
of the effective Particle-Particle interaction. 
In Sec.~\ref{sec:numerical}, we describe the simulation setup and 
discuss the results.  
Finally, we summarize our outcomes in Sec.~\ref{sec:discuss}.

\section{Methodology}\label{sec:method}
\subsection{Schr{\"o}dinger-Poisson equations}
The nature of FDM can be well described by the Schr{\"o}dinger-Poisson equations, 
\begin{equation}\label{schrodinger}
i\hbar\dfrac{d\Psi}{dt}=-\dfrac{\hbar^2}{2m_\chi}\boldsymbol{\nabla}^2\Psi+m_\chi V\Psi, 
\end{equation}
and
\begin{equation}\label{poisson}
\boldsymbol{\nabla}^2V=4\pi Gm_\chi|\Psi|^2. 
\end{equation}
Here $\hbar$, $m_\chi$ and $V$ are 
the Planck constant, particle mass and the gravitational potential acting on a particle, respectively.   
The wave function $\Psi$ can be written as 
\footnote{The original FDM paper~\cite{hu2000fuzzy} has a factor $1/2$ in the wave function 
for identical dark matter particles. 
However, this normalization factor will not change the Lagrangian density  
and the size of quantum pressure in our study. }
\begin{equation}\label{psi}
\Psi=\sqrt{\dfrac{\rho}{m_\chi}}\exp(\dfrac{iS}{\hbar})
\end{equation}
in terms of the number density $\dfrac{\rho}{m_\chi}$, 
while we can define the gradient of $S$ to be the DM momentum,
\begin{equation}\label{velocity}
\boldsymbol{\nabla} S=m_\chi\boldsymbol{v}. 
\end{equation}
After solving the Schr{\"o}dinger-Poisson equations, from the real and imaginary parts of the solution, 
one can obtain the continuity equation, 
\begin{equation}\label{continuity}
\dfrac{d\rho}{dt}+\boldsymbol{\nabla}\boldsymbol{\cdot}(\rho \boldsymbol{v})=0, 
\end{equation}
and the momentum-conservation equation, 
\begin{equation}\label{momentum}
\dfrac{d\boldsymbol{v}}{dt}+(\boldsymbol{v}\boldsymbol{\cdot}\boldsymbol{\nabla})\boldsymbol{v}=-\boldsymbol{\nabla}(Q+V), 
\end{equation}
where we have defined the quantum pressure as
\begin{equation}\label{pressure}
Q=-\dfrac{\hbar^2}{2m_\chi^2}\dfrac{\boldsymbol{\nabla}^2\sqrt{\rho}}{\sqrt{\rho}}.
\end{equation}
Eqs.~\eqref{continuity} and ~\eqref{momentum} are known as Madelung equations
~\citep{spiegel1980fluid,Uhlemann:2014npa,Marsh:2015daa}.
One can see that such a pressure is only related to the mass density $\rho$ and 
can be treated as a new force on the particles additional to gravity. 
Later, we shall focus on this pressure term and discuss how to obtain 
the acceleration information by using the Hamiltonian field theory.

To discuss the effect of the quantum pressure, we can start with the Hamiltonian without the gravity term,  
\begin{equation}\label{hamiltonian}
H=\int\dfrac{\hbar^2}{2m_\chi}|\boldsymbol{\nabla}\Psi|^2d^3x=\int\dfrac{\rho}{2}|\boldsymbol{v}|^2d^3x+
\int\dfrac{\hbar^2}{2m_\chi^2}(\boldsymbol{\nabla}\sqrt{\rho})^2d^3x. 
\end{equation}
We can write the kinetic energy term in discretized form 
with particle index $j$ 
\begin{equation}\label{kinetic}
T=\int\dfrac{\rho}{2}|\boldsymbol{v}|^2d^3x=\sum_j\dfrac{1}{2}m_j(\dfrac{dq_j}{dt})^2,
\end{equation}
where $q_j$ is the coordinate of the $j$th particle, 
and the effective potential energy is from the quantum pressure
\begin{equation}\label{eq:potential}
K_\rho=\int\dfrac{\hbar^2}{2m_\chi^2}(\boldsymbol{\nabla}\sqrt{\rho})^2d^3x. 
\end{equation}
Note that we did not discretize $K_\rho$ here 
but delay it to the next subsection
because it will require some efforts to do so. 
Based on $T$ and $K_\rho$ the Lagrangian of the system without gravity is 
\begin{equation}\label{lagrangian}
L=T-K_\rho=\sum_j\dfrac{1}{2}m_j(\dfrac{dq_j}{dt})^2-\int\dfrac{\hbar^2}{2m_\chi^2}(\boldsymbol{\nabla}\sqrt{\rho})^2d^3x,  
\end{equation}
and the Euler-Lagrangian equation becomes 
\begin{equation}\label{euler-lagrangian}
\dfrac{d}{dt}\dfrac{\partial L}{\partial \dot{q_j}}-\dfrac{\partial L}{\partial q_j}=0
\Longrightarrow m_j\ddot{q_j}=-\dfrac{\partial K_\rho}{\partial q_j}.
\end{equation}
One can see that the $\rho$ in $K_\rho$ is a continuous function, which cannot
be used 
in the PP method. Therefore, the major task is to further discretize 
the continuous function $\partial K_\rho/\partial q_j$,
which we shall describe in more details in the next subsection.
We clarify here that, the index $i$ and $j$ here are representing the simulation particles
rather than single "Ultra-Light Axion" particles. 

\subsection{Particle-particle implementation of quantum pressure}
For a particle-particle interaction system, 
the number density for each individual particle is a delta function.  
Intuitively, the mass density $\rho$ can be discretized as 
\begin{equation}\label{eq:rho}
\rho(\boldsymbol{r})=\sum_i m_i\delta(\boldsymbol{r}-\boldsymbol{r_i}), 
\end{equation}
where the summation over the index $i$ means to add up all the particles.
Numerically, treatment of a delta function is a difficult computational problem 
because of the sampling coverage issue. However, conventionally one can approximate a delta function 
as a narrow Gaussian/kernel function, as long as the width is small enough. 
The reasons and advantages to use the Gaussian smoothing kernel are listed as follows.  
(i) It naturally keeps the kernel smooth, differentiable, and spherically symmetric.
(ii) The particle-particle interaction can naturally avoid the singularity at zero-density positions. 
Because of the finite grid size, such a singularity can numerically make the results unphysical.
Specifically, we write down the form of the delta function as
\begin{equation}\label{eq:delta_function}
\delta(\boldsymbol{r}-\boldsymbol{r_i})=\dfrac{2\sqrt{2}}{\lambda^3\pi^{3/2}}\exp(-\dfrac{2|\boldsymbol{r}-\boldsymbol{r_i}|^2}{\lambda^2}),
\end{equation}
with a narrow width $\lambda$. Note that the value of $\lambda$ is not arbitrary and should be 
the same as the de Broglie wavelength because one FDM particle has to be found with a
high probability within a Gaussian wave packet, 
motivated by the wave function Eq.~\eqref{psi}.
In our work, the probability of finding a FDM in one wavelength is set at $95\%$.

Taking the FDM mass around $\mathcal{O}(10^{-31})\gev$ as an example, its wavelength $\lambda$ is order of kpc. Inserting Eq.~(\ref{eq:delta_function}) back to 
the $(\boldsymbol{\nabla}\sqrt{\rho})^2$ term in Eq.~(\ref{eq:potential}), 
the expression can be expanded by using the kernel function,    
\begin{eqnarray}\label{eq:nablasqrtrho}
\left[\boldsymbol{\nabla}\sqrt{\rho(\boldsymbol{r})}\right]^2&=&
\dfrac{1}{4\rho(\boldsymbol{r})}\left[\sum_i m_i\boldsymbol{\nabla}\delta(\boldsymbol{r}-\boldsymbol{r_i})\right ]^2, \nonumber \\
&=&\dfrac{1}{4\rho(\boldsymbol{r})}\left[\sum_i m_i\delta(\boldsymbol{r}-\boldsymbol{r_i})(-\dfrac{4}{\lambda^2})(\boldsymbol{r}-\boldsymbol{r_i})
\right ]^2, \\
&=&\dfrac{4}{\lambda^4\rho(\boldsymbol{r})}\left[\sum_i m_i\delta(\boldsymbol{r}-\boldsymbol{r_i})(\boldsymbol{r}-\boldsymbol{r_i})\right]^2.\nonumber
\end{eqnarray}
In the simulation, those FDM particles such as axions will be grouped into a big mass clump in space 
which can be treated as an imaginary particle point 
(neglecting the size of the clump in the cosmological scale), and the mass density, Eq.~(\ref{eq:rho}), becomes 
\begin{equation}\label{eq:rho_group}
\rho(\boldsymbol{r})=\sum_j \sum_i m_i\delta(\boldsymbol{r}-\boldsymbol{r_j}), 
\end{equation} 
with the index $j$ for each imaginary particle clump. 
Mathematically, one can think that the mass density is expanded around $\boldsymbol{r_j}$ 
to include all the FDM particles, $\boldsymbol{r}\to\boldsymbol{r}-\boldsymbol{r_j}$ and 
$\boldsymbol{r_i}\to\boldsymbol{r_i}-\boldsymbol{r_j}$.   
Given such a consideration, 
the summation of individual FDM particles is effectively the same as summing over 
all the imaginary particle points, and Eq.~(\ref{eq:nablasqrtrho}) can be further polished to be  
\begin{equation}
\label{eq:group_particle}
\left[\boldsymbol{\nabla}\sqrt{\rho(\boldsymbol{r})}\right]^2
\simeq\dfrac{4}{\lambda^4}
\left[\sum_j m_j\delta(\boldsymbol{r}-\boldsymbol{r_j})(\boldsymbol{r}-\boldsymbol{r_j})\right]^2
\left[\sum_j m_j\delta(\boldsymbol{r}-\boldsymbol{r_j})\right]^{-1}. 
\end{equation}
It is worth mentioning that the form in Eq.~(\ref{eq:group_particle}) is identical to Eq.~(\ref{eq:nablasqrtrho}) 
but their meanings should not be confused. Hence, we leave the different indices here.    
At this stage, we have successfully converted the Gaussian wave packet 
into an imaginary particle-smoothing kernel.

To completely discretize $\partial K_\rho/\partial q_j$, we still need to integrate 
Eq.~(\ref{eq:group_particle}) over all space. 
Due to the nature of the delta function, we just need to focus on the volume surrounding 
the imaginary particle points. Therefore, the integration together with kernel approach gives
\begin{eqnarray}
\int \left[\boldsymbol{\nabla}\sqrt{\rho(\boldsymbol{r})}\right]^2 dx^3
&\simeq&\int \dfrac{4dx^3}{\lambda^4}
\left[\sum_j m_j\delta(\boldsymbol{r}-\boldsymbol{r_j})(\boldsymbol{r}-\boldsymbol{r_j})\right]^2
\left[\sum_j m_j\delta(\boldsymbol{r}-\boldsymbol{r_j})\right]^{-1}  \label{eq:18}\\
&\simeq &4\lambda^{-4}\sum_j m_j\delta(\boldsymbol{r}-\boldsymbol{r_j})(\boldsymbol{r}-\boldsymbol{r_j})^2\Delta V_j \mathcal{B}_j \label{eq:19}\\
&\simeq &4\lambda^{-4}\sum_j m_j\frac{\Delta V_j\mathcal{B}_j}{\lambda^3\pi^{3/2}} \exp\left[-\dfrac{(\boldsymbol{r}-\boldsymbol{r_j})^2}{\lambda^2}\right]
(\boldsymbol{r}-\boldsymbol{r_j})^2,
\end{eqnarray}
where the parameters $\Delta V_j$ and $\mathcal{B}_j$ are the effective volume 
and correction factor of the $j$th simulation particle which will be described 
in more detail below.  

We propose a correction factor $\mathcal{B}_j$ for the $j$th simulation particle 
in order to numerically take care of the different integration results  
between the delta function and Gaussian kernel. 
In other words, when we treat a delta function as 
one Gaussian kernel with width equal to one matter wavelength, 
it does not behave like a delta function in the region 
where the distance between two kernel centers is less than one wavelength. 
In such a short range, the overlap between 
two Gaussian tails can 
also contribute significantly, especially when performing integration with high particle density. 
For more detailed explanations and the fitting formula for $B_j$ Eq.~\eqref{BiasFit},  
see Appendix~\ref{app:Bfactor}.

Theoretically, the effective volume $\Delta V_j$ for each simulation particle $j$ is of
the order of $\lambda^3\pi^{3/2}$ resulting from a Gaussian kernel integral.  
However, the exact value of $\Delta V_j$ can differ from system to system  
because of the complexity of the inner kernel region.   
Hence, we treat it as a phenomenological free parameter (a constant for simplicity), 
but we adjust its value to match the result inside the soliton 
core obtained by other approaches 
which have better resolution in the region
less than one wavelength, such as Ref.~\cite{schive2014cosmic}.

Finally, Eq.~\ref{eq:potential} can be simply rearranged as 
\begin{equation}\label{partialqi}
\sum_j\frac{\partial K_\rho}{\partial q_j}
=\dfrac{4\hbar^2}{m_\chi^2\lambda^4}\sum_j m_j\Delta V_j \mathcal{B}_j
\exp\left[-\dfrac{2|\boldsymbol{r}-\boldsymbol{r_j}|^2}{\lambda^2}\right]
(1-\dfrac{2|\boldsymbol{r}-\boldsymbol{r_j}|^2}{\lambda^2})(\boldsymbol{r}-\boldsymbol{r_j}),
\end{equation}
and the equation of motion, Eq.~(\ref{euler-lagrangian}), becomes 
\begin{equation}\label{final}
\sum_j m_j \ddot{q}=-\dfrac{4\hbar^2}{m_\chi^2\lambda^4}\sum_j m_j\Delta V_j \mathcal{B}_j 
\exp\left[-\dfrac{2|\boldsymbol{r}-\boldsymbol{r_j}|^2}{\lambda^2}\right]
(1-\dfrac{2|\boldsymbol{r}-\boldsymbol{r_j}|^2}{\lambda^2})(\boldsymbol{r}-\boldsymbol{r_j}).
\end{equation}
Substituting $q$ with $\boldsymbol{r}$, the additional acceleration from quantum pressure 
used in the simulation can be written as 
\begin{equation}\label{eq:accel}
\ddot{\boldsymbol{r}}=\dfrac{4M\hbar^2}{M_0 m_\chi^2 \lambda^4}\sum_j\mathcal{B}_j\exp
\left[-\dfrac{2|\boldsymbol{r}-\boldsymbol{r_j}|^2}{\lambda^2}\right]
(1-\dfrac{2|\boldsymbol{r}-\boldsymbol{r_j}|^2}{\lambda^2})(\boldsymbol{r_j}-\boldsymbol{r}).
\end{equation} 
Here $M$ is the mass of the simulation particle and $M_0$ is a normalization factor 
accounting for the size of $\Delta V_j$, which we choose to be $10^6 M_{\odot}$.
Interestingly, if we put any ``one" test particle around some 
quantum pressure sources, the additional energy to the system 
injected from quantum pressure term is zero. 
Namely, the total work, the integration of Eq.~(\ref{eq:accel}) 
from $r=0$ to $r=\infty$, vanishes.

\begin{figure}
\includegraphics[scale=0.45]{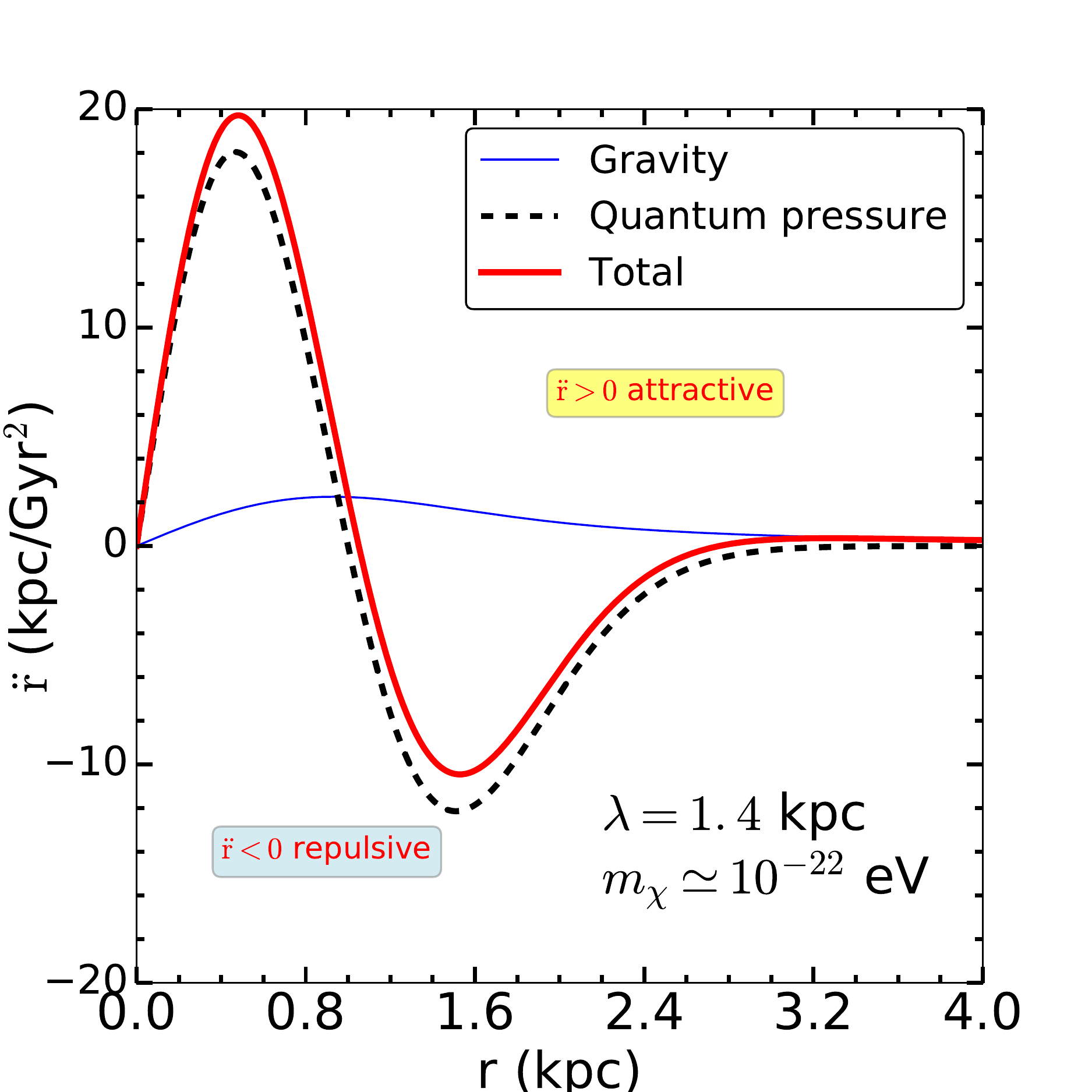}
\includegraphics[scale=0.45]{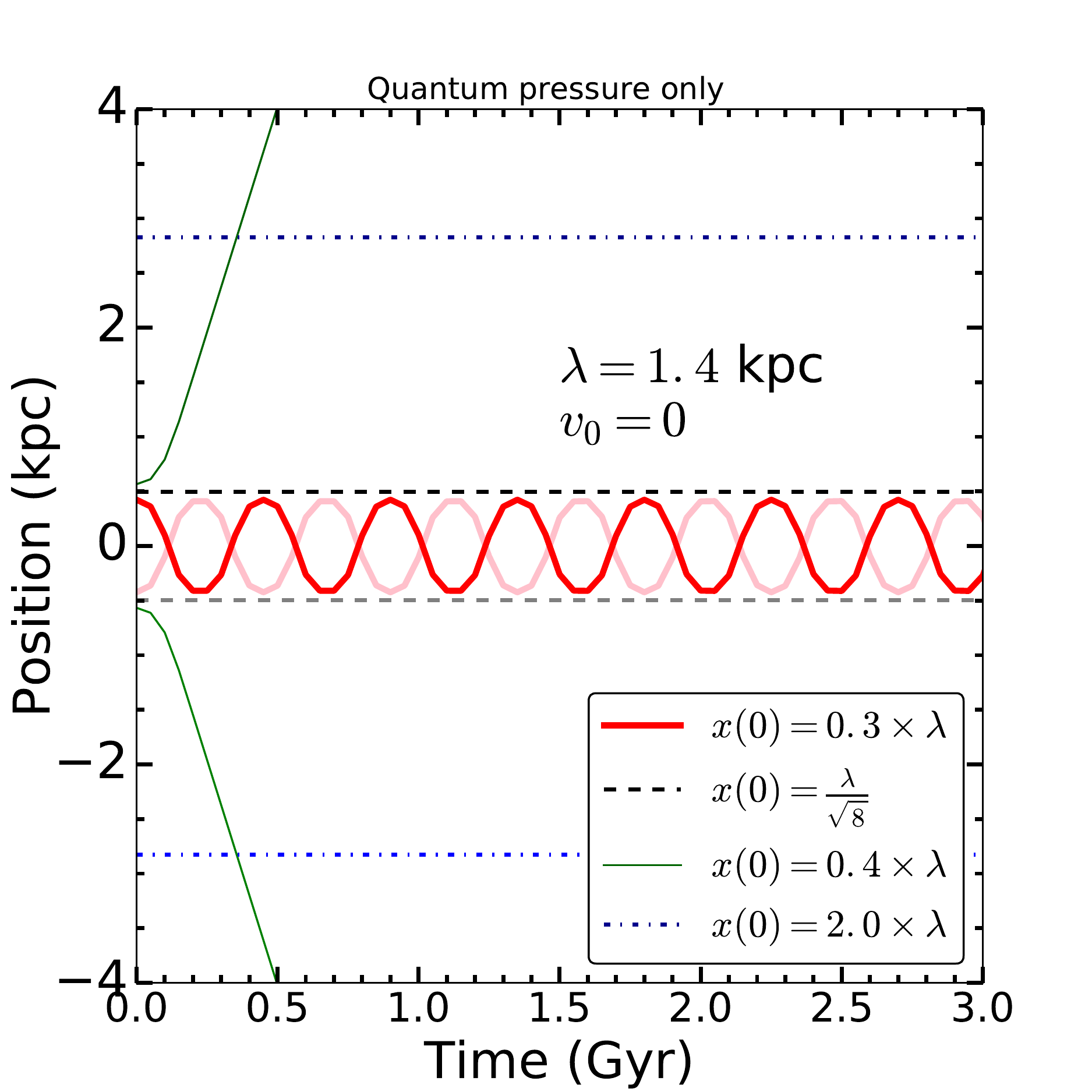}
\caption{Left panel: the acceleration from 
quantum pressure (black dashed), gravity (blue solid), and their sum (red solid) between two particles. 
The x-axis is the distance between two particles while the y-axis is the acceleration.
Right panel: the particle position, which is changing due to the quantum pressure, vs. time. 
The red-solid, black-dashed, green-solid, and blue-dash-dotted lines represent the solutions with initial positions 
at $0.3\lambda$, $\lambda/\sqrt{8}$, $0.4\lambda$, and $2\lambda$, respectively. 
However, the light color lines represent their partner particles located at 
the opposite side $-0.3\lambda$, $-\lambda/\sqrt{8}$, $-0.4\lambda$, and $-2\lambda$ initially.  
\label{Fig:pressurefigure}}
\end{figure}

To illustrate the effect of quantum pressure, let us consider 
a two-particle system ($\mathcal{B}_j=1$) 
separated by a distance of order $\mathcal{O}(\kpc)$ and the 
acceleration caused by quantum pressure will be 
$\mathcal{O}(\dfrac{\hbar^2}{m^2 \lambda^3})\sim\mathcal{O}(10^{-10} m/s^2)$. 
In the left panel of Fig.~\ref{Fig:pressurefigure}, we demonstrate the effect of quantum pressure 
in the plane of ($r$, $\ddot{r}$). The acceleration from 
quantum pressure, gravity, and the sum are shown by the black dashed line, blue line, and red line, respectively. 
Clearly, the quantum pressure can be attractive (positive sign in our definition)
\footnote{
Here the terminology ``attractive'' and ``repulsive'' are just simple phrases
to describe the quantum pressure being positive or negative, respectively.}
if the distance between two particles is less than $\lambda/\sqrt{2}$. 
However, it becomes repulsive (negative sign in our definition) 
if the distance is greater than $\lambda/\sqrt{2}$. 
To understand this, we refer back to the quantum pressure definition in 
Eq.~(\ref{pressure}). 
The pressure term $Q$ is proportional to the second derivative of the mass density, 
namely, the curvature of the density, which can have negative, positive, 
or zero values, physically corresponding to 
attractive, repulsive, and zero forces.

In the right panel of Fig.~\ref{Fig:pressurefigure}, we show the 
position $x(t)$ of one of the particles
in this two-particle system as a function of time by solving 
Eq.~(\ref{eq:accel}). 
Here the origin is located at the center of mass and such that $r(t)=2x(t)$. 
The red-solid, black-dashed, green-solid, and blue-dash-dotted lines present 
the cases with 
initial positions $x(0)$ 
at $0.3\lambda$, $\lambda/\sqrt{8}$, $0.4\lambda$, and $2\lambda$, respectively.
Also, the position of the other particle is drawn in the
corresponding lighter colors for reference. 
Interestingly, when the distance between the two particles is smaller 
than $\lambda/\sqrt{2}$, 
the attractive force will bound them.
Note that this also demonstrates the phenomenon of Bose-Einstein condensation. 
One shall bear in mind that for fermionic particles such short range attractive forces 
would not exist because of the Pauli's Exclusion Principle.
However, the repulsive force will push these two particles away for 
$\dfrac{\lambda}{\sqrt{2}}<r(0)<\sqrt{2}\lambda$. 
At $r(0)=\lambda/\sqrt{2}$, the node represents zero interaction. 
For the $x(0)\gg \sqrt{2} \lambda$ case, the two particles barely feel 
the force from each other 
so that the they will continue with their initial velocities, though we set them to be zero. 

This reveals the quantum pressure as a short-range interaction, shown in Eq.~(\ref{eq:accel}),
with the exponentially decaying term.

%
%
\section{Numerical result}
\label{sec:numerical}
%
%
\texttt{Gadget2}~\citep{gadget2} is a \texttt{TreePM} hybrid N-body code. 
However, in order to describe continuous quantum pressure, 
the interaction term has to be discretized as we have discussed in the 
previous section.   
Therefore, we modified \texttt{Gadget2} to compute the contribution from the quantum pressure. 
In this section, we first describe the details of what we used in our 
simulation, and then we present our simulation results.
   
\subsection{The simulation setup}

The PM method is mostly useful for cosmological simulations with periodic boundary conditions.
As we have seen from the previous section that the quantum pressure behaves like a short-range interaction, 
we therefore keep the original PM code which takes care of the long-range force calculation. 
However, we have to use the Tree method to take care of the short-range force calculation, and so the part of Tree force calculation in \texttt{Gadget2} has to be modified to include 
the quantum pressure term. 
Finally, we do not need to set softening for
the quantum pressure acceleration since it is finite in the $(\boldsymbol{r_j}-\boldsymbol{r})\simeq 0$ region.

To study the differences between the CDM and FDM, we set up a 
self-collapsing system.
A cubic box with a side length $400\kpc$ is generated with 
totally $10^6$ simulation particles homogeneously distributed inside the box with the vacuum boundary condition. The box can be understood as an over-density region which will collapse
under its self-gravity. 
Each simulation particle consists of a $10^6$ solar mass.  
All particles start from rest and the system collapses resulting from their self-gravity to 
form a stable self-gravitationally-bound virialized halo at the center. 
It is sufficient to stop the simulation around $10\gyr$.
The final stable virialized halo in FDM model depends sensitively on the initial slight perturbation given to the system.

Two kinds of \texttt{Gadget2} simulations are performed: 
one with the modified code for FDM quantum pressure effect 
and the other one with the original \texttt{Gadget2} code. 
In both simulations, we use the same initial conditions, and 
the gravitational softening length is chosen to be $0.89\kpc$. 
The ULA mass and wavelength are fixed to be $2.5\times 10^{-22}\ev$ 
and $1.4\kpc$, respectively.
We further run several simulations with different softening lengths in order to make sure that the final density and velocity distributions of the 
halos have converged.

\subsection{Simulation results}

\begin{figure}
\includegraphics[scale=0.25]{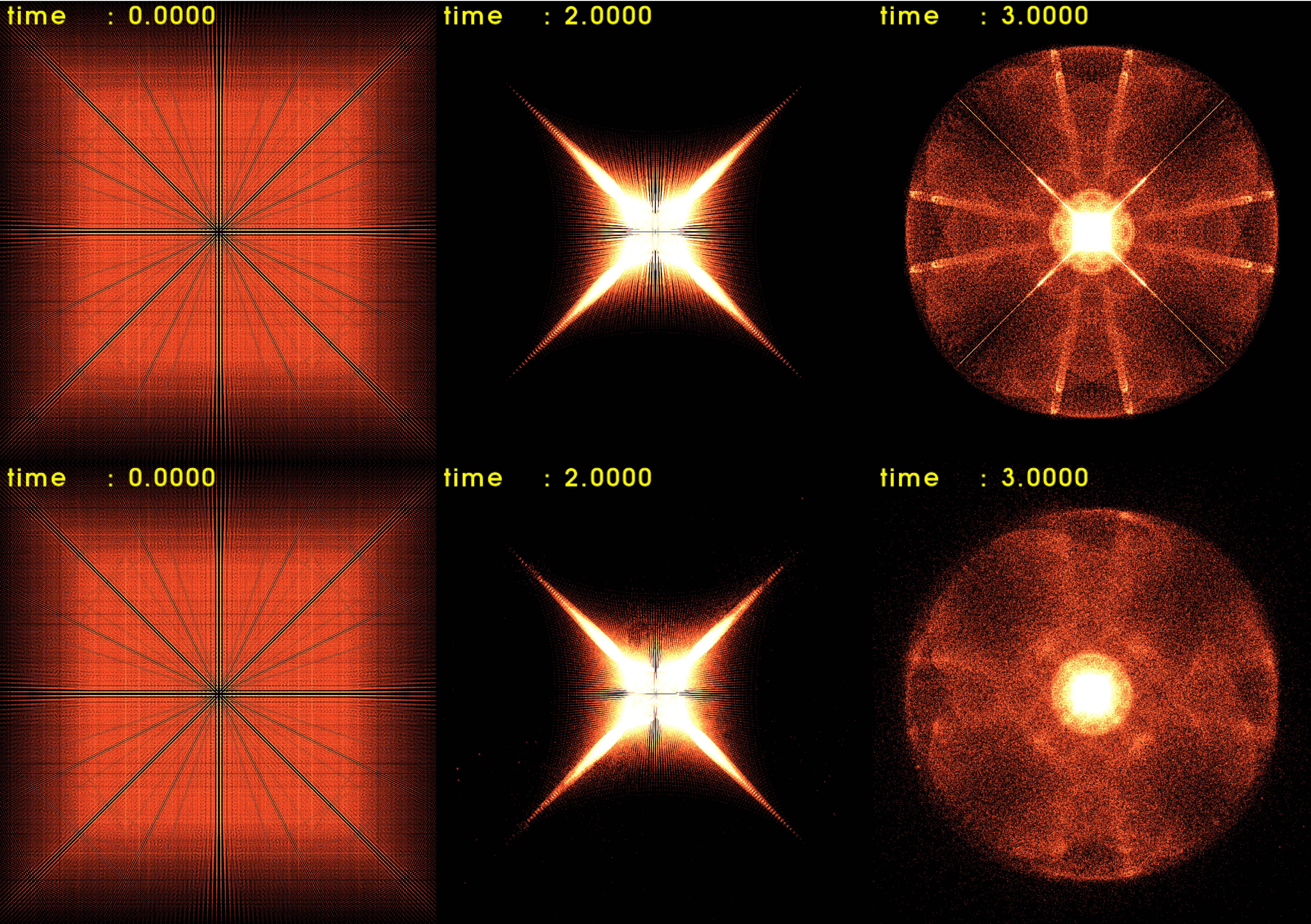}
\caption{
\label{fig:collaps}
The particle distributions of the CDM (on the top) and FDM (on the bottom) 
simulations for a self-gravitationally-collapse system. The labels at the upper left corner show the time in \gyr.}
\end{figure}

In Fig.~\ref{fig:collaps}, we present the particle distributions of our simulations 
for a self-gravitationally-collapse system. 
The three panels on the upper (lower) row are for the CDM (FDM) scenario. The figures from the left to right panels are two-dimensional 
slices at the evolution time $0$, $2$, and $3\gyr$. Clearly, 
the distribution of FDM is more smooth and spread out than the CDM case, especially 
at later time. This is 
exactly the novel feature of the quantum pressure.

\begin{figure}
\includegraphics[scale=0.45]{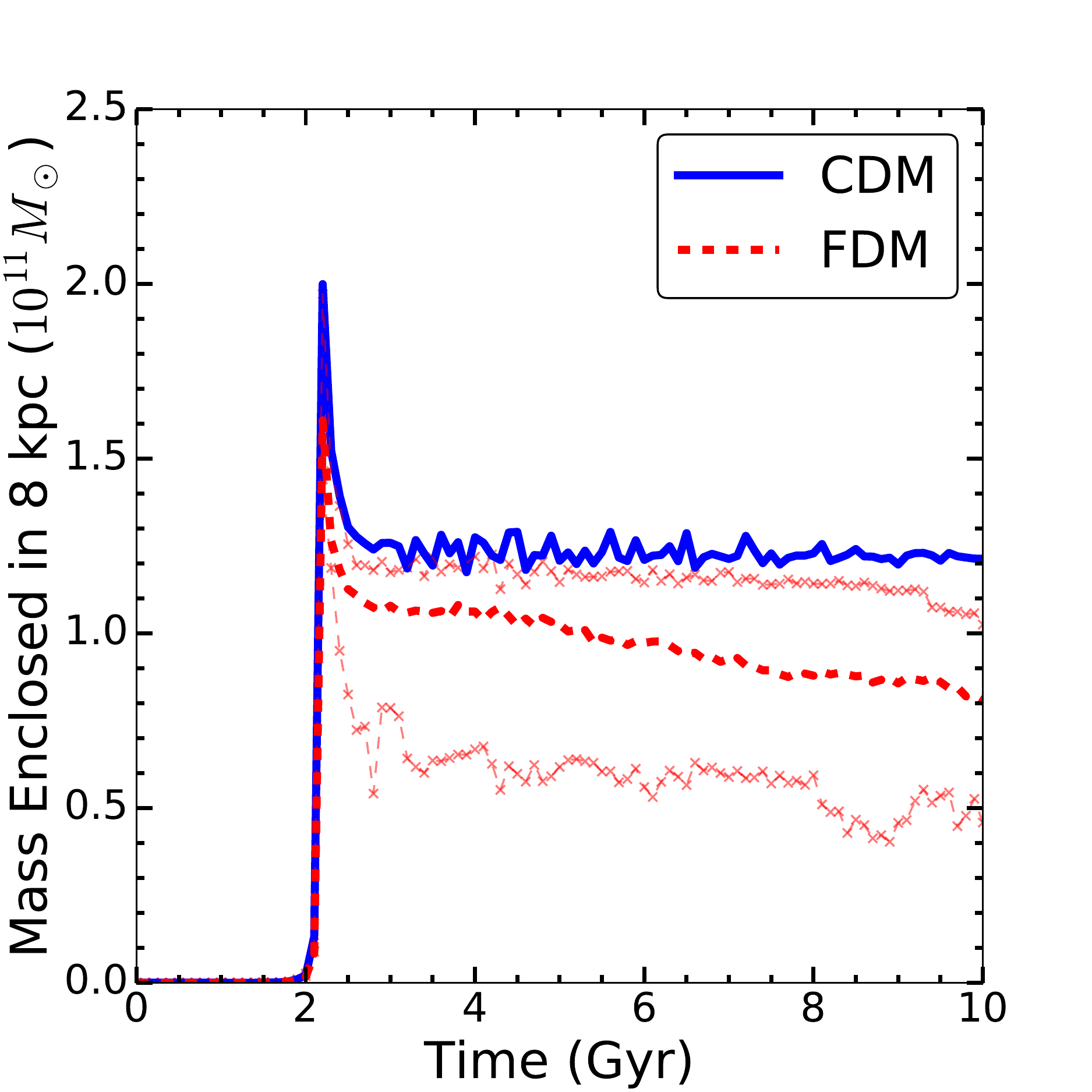}
\includegraphics[scale=0.45]{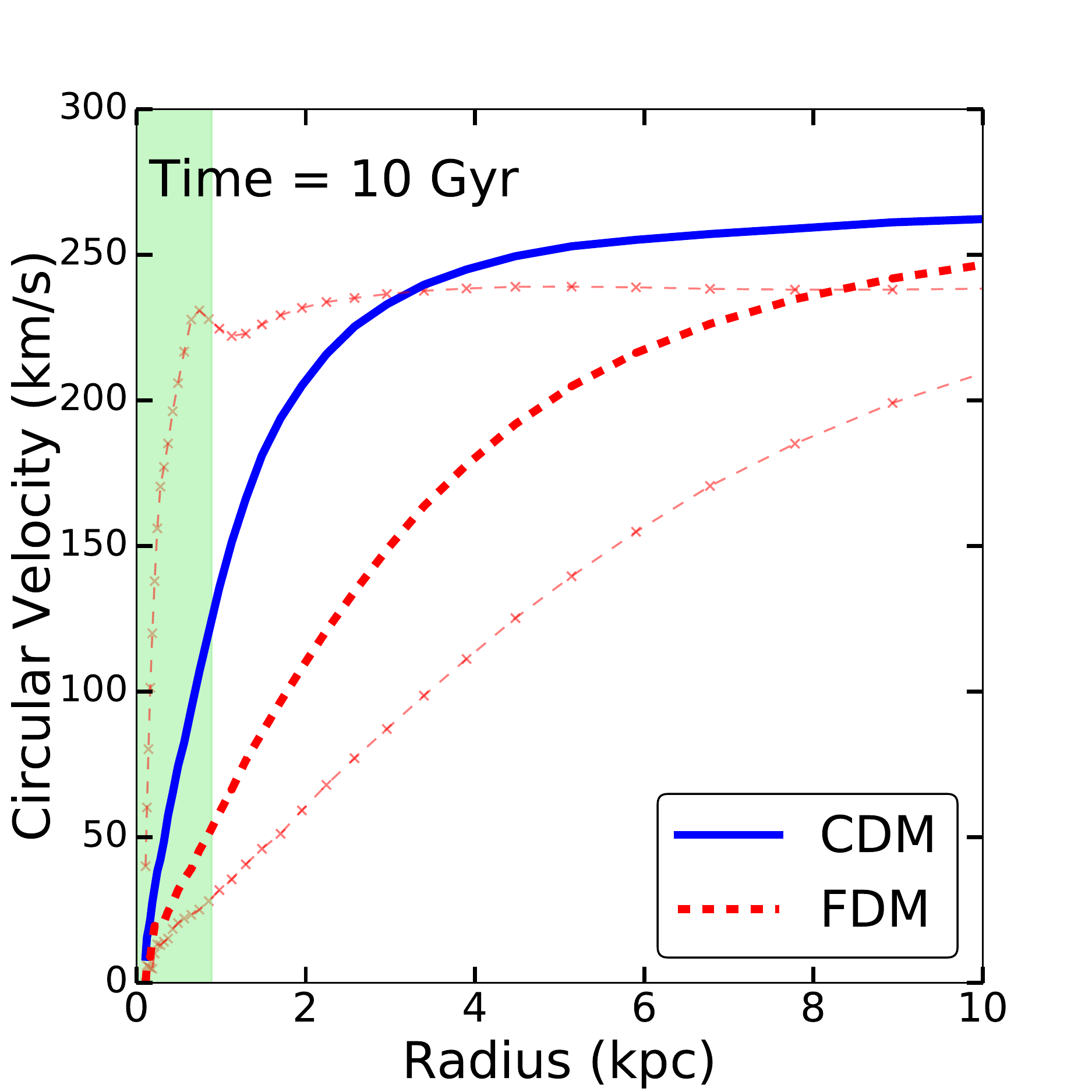}
\caption{
The evolution of mass enclosed in $8\kpc$ from the center is shown on the left
panel.
After $3-4\gyr$ evolution, the self-collapse system reaches equilibrium for both CDM and FDM models. 
The right panel shows the final rotational curve. 
The circular velocity for the FDM halo is smaller than 
the CDM inside $10\kpc$ from the center
Two solutions of FDM without corrections 
(pink thin solid lines) are shown for reference. 
The green shaded region represents the radius of halo smaller than the softening length $0.89\kpc$.
\label{fig:coreMass}}
\end{figure}

To further illustrate the effect of the quantum pressure on the halo density profile, 
we plot the mass enclosed in $8\kpc$ 
in the left panel of Fig.~\ref{fig:coreMass}, 
which shows the core mass evolution 
along the time direction in unit of $\gyr$
for the CDM (blue line) and FDM (red solid line) cases. 
For reference we also plot two curves colored in pink (thin dashed) for $\mathcal{B}_j=1$ case. 
\footnote{Because of the algorithm of the tree method used in
Message Passing Interface (MPI), 
the boundary of the simulation grid can be slightly different. 
Without implementing the correction factor $\mathcal{B}_j$, 
such small differences from Gaussian tails 
can be accumulated and developed into
two different but stable results by using several different node assignments.
} 
Both systems of CDM and FDM will be in the equilibrium state 
after $3-4\gyr$ when the mass within $8 \kpc$ does not change 
significantly.
Note that the FDM halo possesses 
slightly less (about $80\%$ of the) mass inside $8\kpc$ than
the CDM one, due to the quantum pressure. 
The FDM halo is also slowly losing mass inside $8\kpc$ 
even after $4\gyr$ while the CDM is already stable.
However both FDM halo and CDM halo are fully virialized after $4\gyr$,
their overall density profiles are stable. 

In the right panel of Fig.~\ref{fig:coreMass}, 
we show the rotational curves for the FDM 
(red solid line for FDM including correction $\mathcal{B}_j$
but pink thin dashed lines for FDM without correction as references) 
and CDM (blue line) halo after $10\gyr$ evolution. 
We take this halo as an example because it is fully virialized. 
We can see that inside $3\kpc$ the circular velocity for 
the FDM halo is always smaller than that of the CDM halo.
This implies that 
the FDM halo has a lower density in the inner core than that of the CDM.
Note that the green shaded region represents
 the softening length $0.89\kpc$ and  
we should treat it as our numerical simulation resolution 
boundary,  inside which we should not trust the results.

\begin{figure}
\includegraphics[scale=0.45]{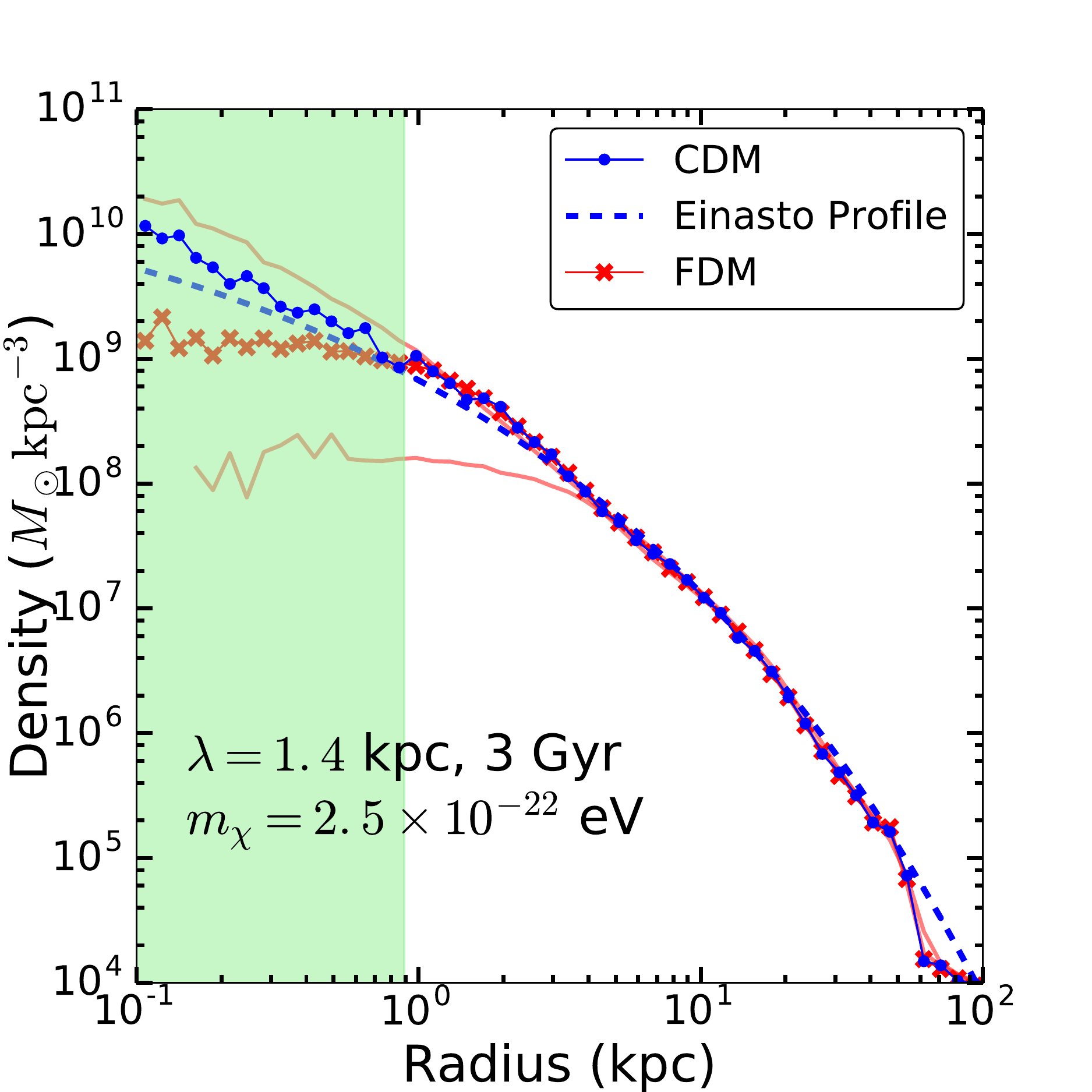}
\includegraphics[scale=0.45]{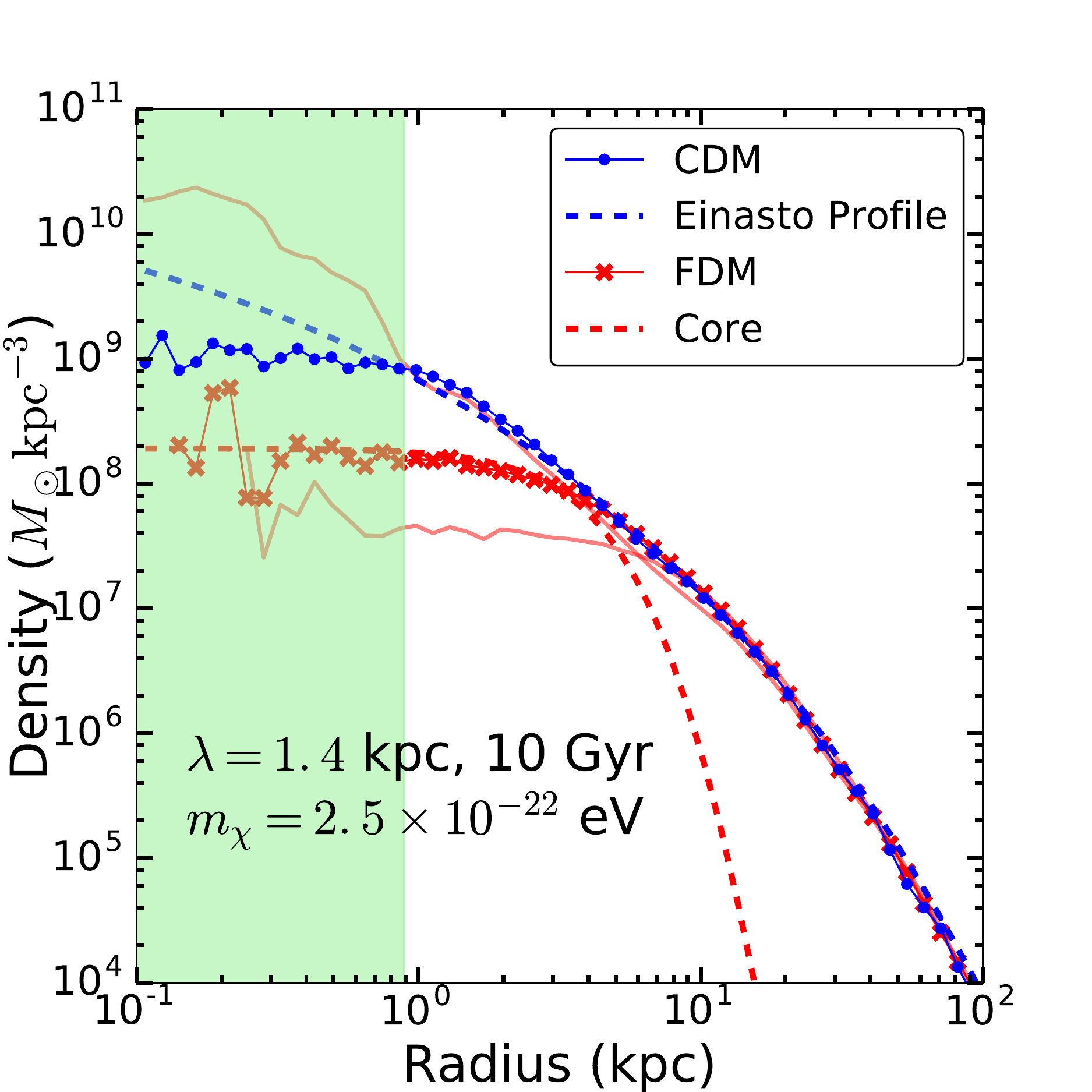}
\caption{The left (right) panel is the halo density profile after $3\, (10)\gyr$. 
The two solutions of FDM without the correction factor $\mathcal{B}_j$
(pink thin solid lines) are shown for reference. 
The green shaded region represents the radius of halo smaller than the softening length $0.89\kpc$.
The system is fully virialized after $10 \gyr$. We can see that the FDM and CDM halo density 
profiles outside $10\kpc$ 
can both be fitted well by the Einasto profile. 
The density inside $1\kpc$ is lower than that of CDM halo after $3 \gyr$ and finally 
evolves into a solitonic core inside $4 \kpc$.
\label{fig:densityProfile}}
\end{figure}

We plot two density profiles at $3\gyr$ and $10\gyr$ separately in Fig.~\ref{fig:densityProfile}. 
In the left panel, we show the density profiles for the FDM (red crosses) and 
CDM (blue circles) halos at $3\gyr$. 
Again, we plot the two solutions of FDM without the correction factor
(pink thin solid lines) as reference. 
The green shaded region represents the region with radius smaller than the 
softening length $0.89\kpc$.
The halo is still not reaching the equilibrium at $3\gyr$ time, and there 
is no significant difference between the CDM and FDM.
We also plot the Einasto Profile fitting in the final virialized state 
in blue dashed line for comparison. 
At the region around $100\kpc$, some particles are still bouncing out so that 
the tail is not fitted well to the Einasto profile.
In the right panel, we show the density profiles at $10\gyr$. 
The fitting function of the solitonic core is shown in red-dashed line,
the formula for which was given in Ref.~\citep{schive2014understanding}, 
\begin{equation}
\rho_c(r)\simeq \rho_b\rho_0[1+0.091(\dfrac{r}{r_c})^2]^{-8}, 
\end{equation}
with an additional fitting parameter $\rho_b$, and 
\begin{equation}
\rho_0\simeq 3.1\times 10^6 (\dfrac{2.5 \times 10^{-22}\ev}{m_{\chi}})^2(\dfrac{\kpc}{r_c})^4\dfrac{M_\odot}{\kpc^3}.
\end{equation} 
We set $\rho_b=5000$ and $r_c=3\kpc$ in our plot.
Beyond the solitonic core, the density profiles of FDM halos are essentially the same as that of CDM and the Einasto profile.

We confirm the existence of a solitonic core, 
slowly emerging from the FDM simulation, with a size of $3\kpc$. 
Inside the solitonic core, our result agrees with
the previous result obtained in 
Ref.~\citep{schive2014understanding}, which 
was based on a grid-based numerical solution of 
the Schr{\"o}dinger-Poisson equations. 
Outside the solitonic core, our result agrees very well with the Einasto profile,
which is also a well-known property of the FDM halo.  
In a previous study~\citep{veltmaat2016cosmological}, a boosted power in 
small scales was also reported for their small-box comoving 
coordinate simulation. 
Comparing to the CDM simulation, Ref.~\citep{veltmaat2016cosmological} found at most $10\%$ more power
in the FDM simulation. 
We can now easily understand such a result by considering 
the quantum pressure in small scales.

\begin{figure}
\includegraphics[scale=0.45]{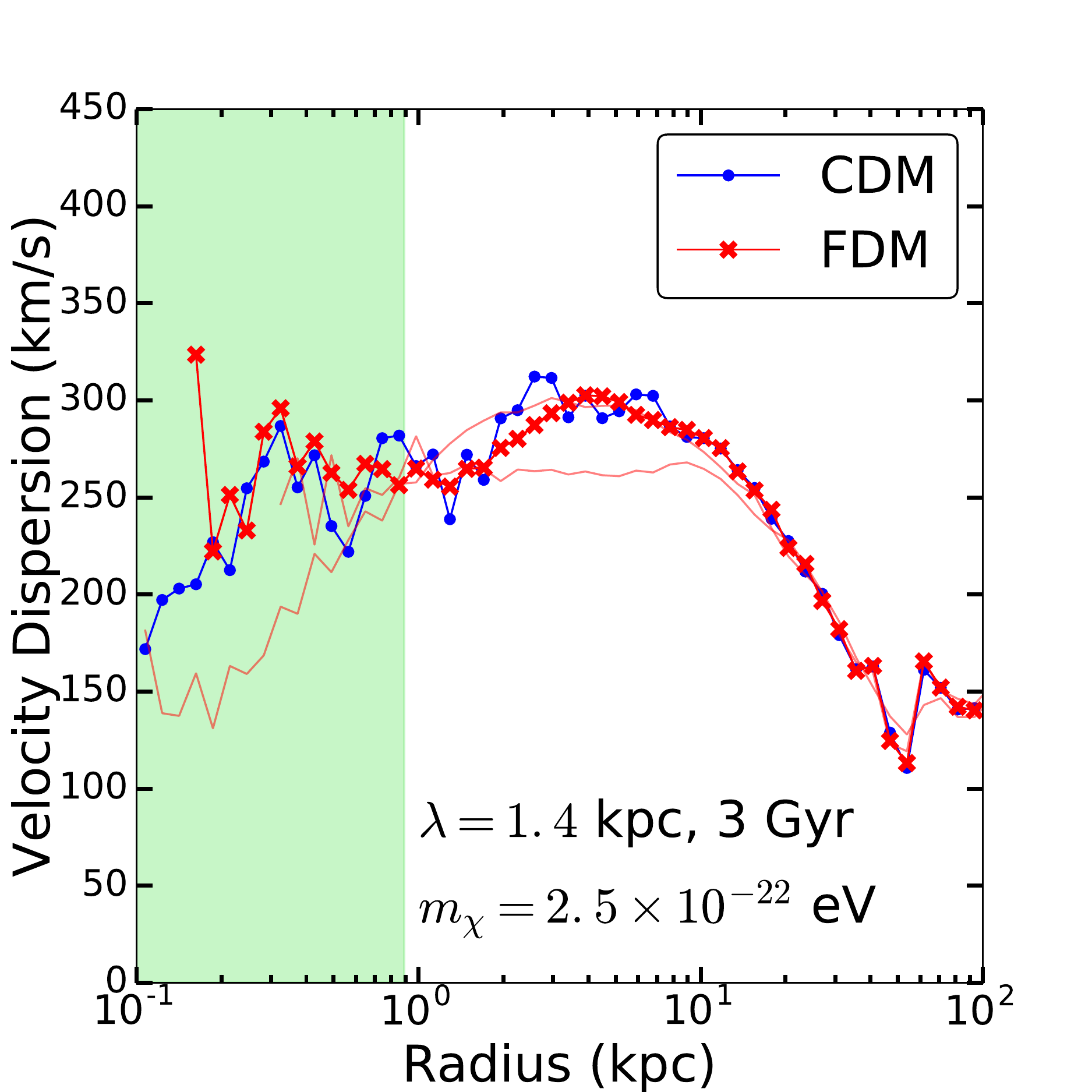}
\includegraphics[scale=0.45]{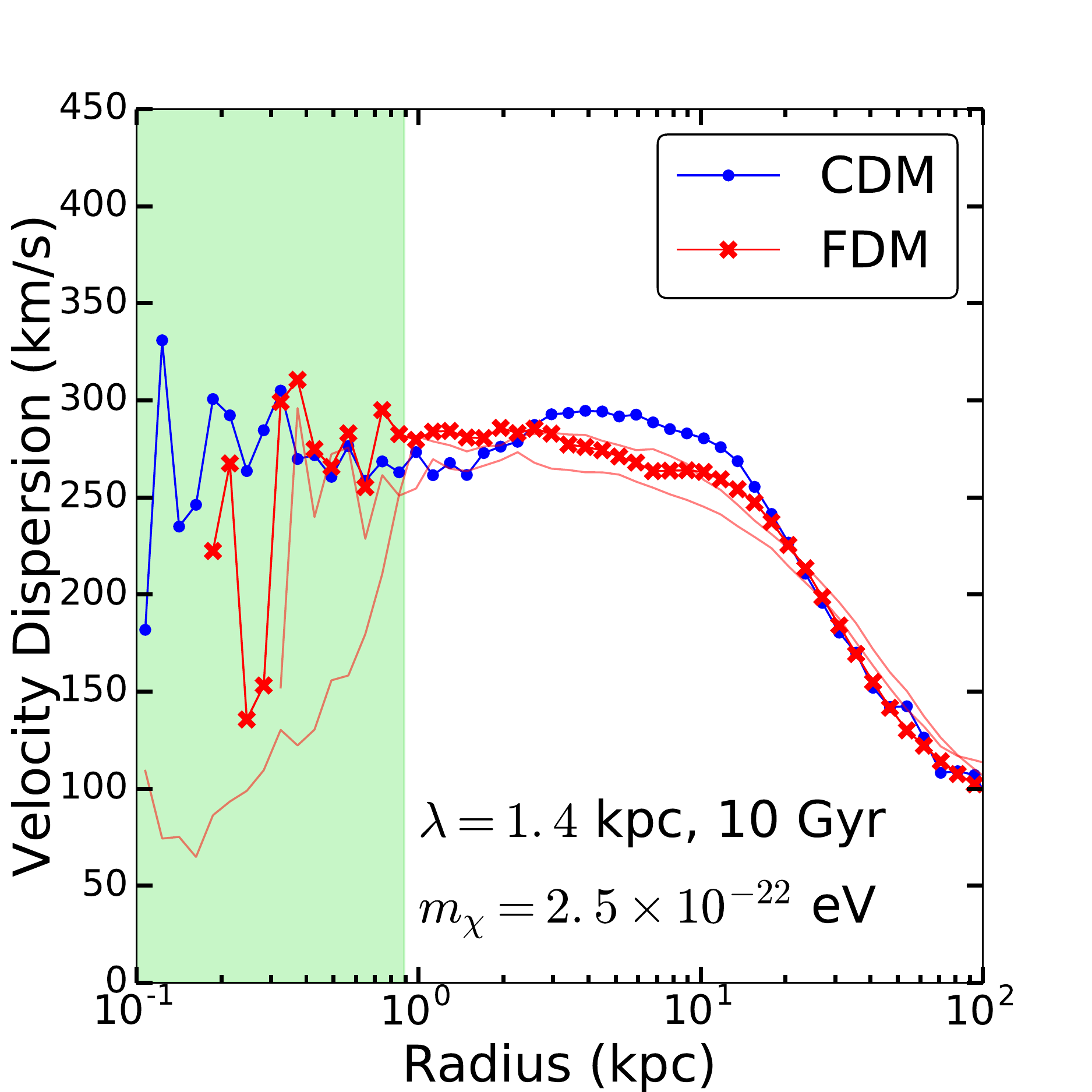}
\caption{Left(right) panel: the velocity dispersion profiles at $3(10)\gyr$. 
The two solutions of FDM without the correction factor $\mathcal{B}_j$
(pink thin solid lines) are shown  for reference. 
The green shaded region represents the radius of halo smaller than
the softening length $0.89\kpc$.
Before final virialization, one cannot find significant difference between the FDM and CDM halos.
The slight differences among the CDM and FDM halos around $10\kpc$ are still clear.
\label{fig:veldispProfile} }
\end{figure}

We also compare the velocity dispersion profiles of the FDM (red lines) 
and CDM (blue line) halos after 3 Gyr (left panel) and after 
10 Gyr (right panel) of evolution in Fig.~\ref{fig:veldispProfile}.
At 3 $\gyr$ (left panel), we do not see significant differences 
between the FDM and CDM halos, 
except in the small central region less than 
the softening length.
However, 
the FDM velocity dispersion is smaller than that of CDM halo between $3\kpc$ and $20\kpc$, 
both at $3 \gyr$ and at  $10\gyr$, which can be understood as due to the lower mass enclosed within
$3\kpc$ in FDM halos, as discussed in Fig.~\ref{fig:coreMass}.

\section{Summary and outlook}
\label{sec:discuss}

In summary, we have proposed to use a Gaussian kernel function to 
discretize the quantum pressure term to simulate FDM in the PP method 
for $N$-body simulations. 
We note that the quantum pressure does not provide additional 
energy to the system, but it will certainly change the halo inner structure. 
In order to understand the quantum pressure effect, we study 
a two-body system and find that the force between the two particles is always 
attractive if the distance between them is less than $\lambda/\sqrt{2}$,  
but it will turn repulsive if the distance between them is larger 
than $\lambda/\sqrt{2}$. 
In small scales, the quantum pressure contribution can be even larger 
than gravity.   

With our discretized quantum pressure approach to the PP method, 
we have constructed two $N$-body simulations, one for the CDM and 
the other one for the FDM, in a collapsing system with 
identical initial conditions.
We found that the FDM halo center can be clearly distinguished from that of 
the CDM and also confirmed that a solitonic core 
forms
 at the FDM halo center, where the mass density is 
very flat, similar to the isothermal or Burkert profiles. 

We have also compared the FDM halo evolution with the CDM case 
based on the mass enclosed within 8 $\kpc$. 
We confirmed that the FDM halo will reach equilibrium 
slightly later than the CDM one while 
its galaxy formation history is consistent with current data~\citep{Hui:2016ltb}.

Implied from the rotation curves, the FDM halo evolves into 
an inner solitonic core with either higher or lower density than the CDM if
they started with identical initial condition; similar results 
can also be found in Ref.~\citep{veltmaat2016cosmological}.

The solitonic core we found in this work
may not be able to quantitatively solve the cusp-core problem,
but it is very suggestive that it may provide a solution 
if a full scale cosmological simulation is performed.
One may worry that 
the non-linear effect from the attractive quantum pressure 
at the region less than one wavelength scale can bring higher 
mass density back to the center. 
This can be understood as the linear power-spectrum growth from the FDM model 
suppresses any power smaller than the FDM Jeans scale  
\begin{equation}
\dfrac{2(\pi G\rho)^{1/4}m_{\chi}^{1/2}}{\hbar^{1/2}}\nonumber \;.
\end{equation}
However, we started our simulations for the CDM and FDM from the 
identical initial conditions, 
and the quantum pressure of the FDM suppresses the density for scales 
larger than one wavelength 
albeit boosts the density up in scales smaller than one
wavelength. 
These two effects were also found in Ref.~\citep{veltmaat2016cosmological} --
the suppression of matter power spectrum from redshift $z=100$ 
to $z=16$ in small scales,
but a boost of matter power spectrum at around $z=2$ in small scales.


We do not consider the cosmological simulation in this work but 
would like to return to this in the near future.

\section*{Acknowledgment}
We would like to thank Volker Springel for some useful suggestion on the code, Tzi-Hong
Chiueh for useful comments on the result.
We acknowledge the help from ITSC in CUHK for providing computational resources for the project.
This project is supported partially by grants from the Research Grant Council of the Hong Kong Special Administrative Region, China (Project No. C4047-14E) and 
the VC Discretionary Fund of CUHK.
K.C. was supported by the MoST under Grants No. MOST-105-2112-M-007-028-MY3.

\appendix

\section{Correction of high-number-density environment}
\label{app:Bfactor}

In the low-density limit, it is reasonable to assume that particles are dominantly under two-body force 
without the contribution from massive overlap terms.
However, in high-number-density environment,
our two-body interaction picture is no longer sufficient because of massive overlap contributions.
In this section, we discuss our strategy to correct this issue.

From Eq.~\eqref{eq:18} and \eqref{eq:19}, if we take $\mathcal{B}_j=1$ we can see that the approximation 
\begin{equation}
\left[\sum_j m_j\delta(\boldsymbol{r}-\boldsymbol{r_j})(\boldsymbol{r}-\boldsymbol{r_j})\right]^2
\left[\sum_j m_j\delta(\boldsymbol{r}-\boldsymbol{r_j})\right]^{-1} \\
\simeq \sum_j m_j\delta(\boldsymbol{r}-\boldsymbol{r_j})(\boldsymbol{r}-\boldsymbol{r_j})^2,
\end{equation} 
will not hold
 when we replace the delta function with the Gaussian kernel. 
Clearly, the approximation only works when the two-body interaction is dominant. 
In reality, a high-number-density system, e.g. in the center of a halo, 
has to be considered a more complicated system.


Let us start with a simple setup.
Assuming $N$ particles occupy a volume $V$, the average number density 
is $N/V$ and the average distance $D$ can be defined as $(N/V)^{1/3}$. 
In the limit of the average distance much larger
than the matter wavelength $D\gg\lambda$, 
the probability for any particle pair to appear within one wavelength  
is given by
\begin{equation}
\mathcal{P}_0\propto (\lambda/D)^3.
\end{equation}
Extending the system to three particles, 
the probability of two particles being within one wavelength
from one another will be proportional to 
$\mathcal{P}_0^2$. 
Similarly, the probability will be proportional to 
$\mathcal{P}_0^{N-1}$ for a $N$-particle system.
Thus, we can conclude that the sum of all pair-interactions can be 
a very good approximation (leading order) for an $N$-body system as long as 
$D\gg\lambda$.  

On the other hand, if we consider the Gaussian kernel $\gau$ 
in the limit of $D\ll\lambda$, 
it is no longer accurate to ignore the overlap between
 the Gaussian tails. 
Such next-to-leading order contributions can be very 
significant if we 
add up all the tail interactions of the $N$-body system.
Hence, we propose a numerical correction factor $\mathcal{B}_j$ to 
account for such additional interactions. 
To estimate the value of $\mathcal{B}_j$ whose index $j$ denotes
the label of tree-node, 
we have to unfold the tree-method back to Eq.~\eqref{eq:rho_group}. 
Therefore, we can first define $\mathcal{B}_j$ (for each $j$) 
to be the net effect of Gaussian tails of all the individual 
simulation particles $i$,
\begin{equation}
\label{eq:bdef}
\left[\sum_i m_i\gau(\boldsymbol{r}-\boldsymbol{r_i})\right]^2
\left[\sum_i m_i\gau\right]^{-1} \\
=\mathcal{B}_j \sum_i m_i\gau(\boldsymbol{r}-\boldsymbol{r_i})^2.
\end{equation} 
We can rewrite the left hand side of Eq.~\eqref{eq:bdef} to be
\begin{eqnarray}
\frac{\sum_{i=1}^{N} m_i^2\mathcal{G}^2(\boldsymbol{r}-\boldsymbol{r_i})
(\boldsymbol{r}-\boldsymbol{r_i})^2
+2\sum_{i\neq i'} m_i m_{i'} 
\mathcal{G}(\boldsymbol{r}-\boldsymbol{r_i})
\mathcal{G}(\boldsymbol{r}-\boldsymbol{r_{i'}})
\left [
\left(\boldsymbol{r}-\boldsymbol{r_i}\right)\cdot
 \left(\boldsymbol{r}-\boldsymbol{r_{i'}}\right )
\right ]
}
{\sum_{i=1}^{N} m_i\mathcal{G}(\boldsymbol{r}-\boldsymbol{r_i}) }. 
\end{eqnarray}
We can see that the major difference between the left hand side (LHS) and right hand side (RHS) of 
Eq.~\eqref{eq:bdef} mainly comes
from the overlap terms, which can be amplified for
$N$ particles. 
Thus, we can guess that the factor $\mathcal{B}_j$ is proportional to $N\sim (D/\lambda)^3$. 
However, finding the exact form for $\mathcal{B}_j$ is not so trivial. 
Here we can perform a Monte-Carlo simulation 
in the range of $D/\lambda$ from zero to infinity
to compute exactly the value of the LHS 
and RHS of Eq.~\eqref{eq:bdef}, and then a proper fitting function
for $\mathcal{B}_j$ can be obtained.
%

In the following, we list the procedures 
of such a Monte-Carlo simulation.
\begin{enumerate}
\item Generate a large number of particles randomly distributed 
in a cubic box. The length of box side is much larger than the matter wavelength $\lambda$.
\item Calculate the 
values of the LHS and RHS of Eq.~\eqref{eq:bdef} 
with $\mathcal{B}_j=1$ at the center of the box.
\item Compare the two values computed in step 2 
and calculate 
the value of $\mathcal{B}_j$ with respect to $D/\lambda$.
\item Repeat the previous steps with different numbers 
of particles, different box sizes
and different $\lambda$'s to ensure the convergence of the 
results.
\end{enumerate}
\begin{figure}[th!]
\centering
\includegraphics[width=3.5in, height=3.5in]{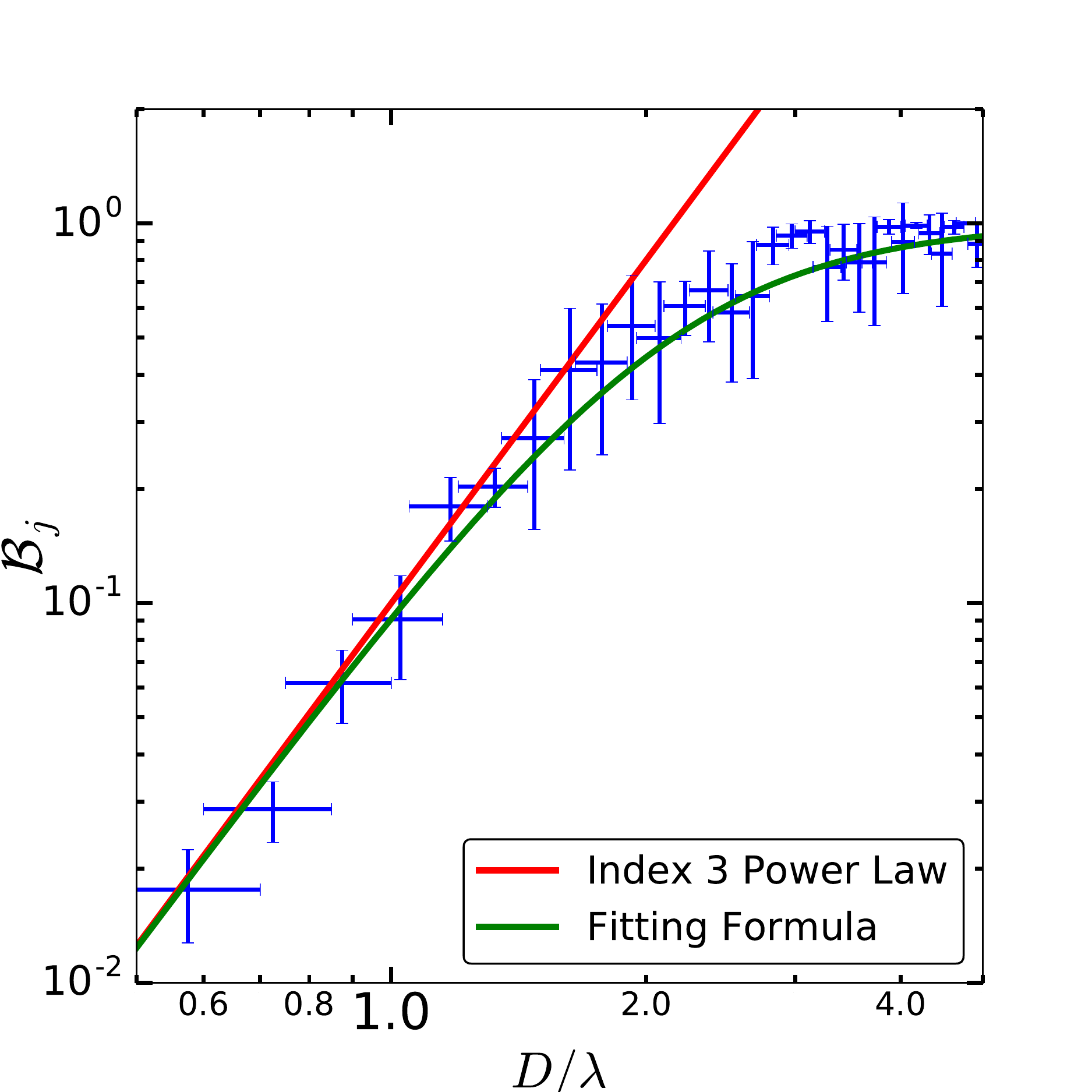}
\caption{\small \label{BiasFigure}
Simulation results are shown in blue errorbars. The red solid line is the power law with power-index=3. The green solid line is the fitting formula we give in \eqref{BiasFit}. The errorbars in the x-axis are given by the bin size. The errorbars in the y-axis are given by the standard deviation of $B_j$ in the bins.}
\end{figure}
We present our simulation results in Fig.~\ref{BiasFigure} and  
fit the data with a formula  
\begin{equation}
\label{BiasFit}
\mathcal{B}_j=(D/\lambda)^3/(10+(D/\lambda)^3).
\end{equation}
The error bars of each data point 
originate from the standard deviation of different scans. 
As expected, the $\mathcal{B}_j$ approaches the value about 
1 at the region of very large $D/\lambda$,
but is almost linearly increasing 
in the log-log scale in the region $D<2\lambda$.
The fitting formula \eqref{BiasFit} provides us the correction of 
the quantum pressure in the densest region in the simulation, 
for example at the center of halos.

\subsection{Numerical implementation of quantum pressure and its corrections in the Tree algorithm}

\texttt{Gadget2}~\citep{gadget2} is a \texttt{TreePM} hybrid N-body code and  
the correction factor 
$\mathcal{B}_j$ has to be implemented in the Tree data structure. 
In the following, we list the basic procedures of the new implementation. 
\begin{enumerate}
\item We build a tree structure to save the information of particle distribution.
\item For each simulation particle, we calculate the gravitational force from the tree nodes.
\item The quantum pressure from the same tree nodes are also computed in the same time and same numerical loop.
\item The local particle number density information can be taken from the tree so that 
the correction factors $\mathcal{B}_j$ can be calculated accordingly.
\item To avoid breaking Newton's third law, 
we also take the particle number density for the father
node of the particle into account. 
Such an action properly keeps the density around the particle
and the tree node smooth without sharp artificial cuts,
which could break the Newton's third law. 
\end{enumerate}  
In principle, the computational time for a FDM simulation is about twice 
as much as that for a classical
gravity only CDM simulation. 
In fact, among our tests the FDM simulation costs two to three
times as much as the CDM one. 
When particles are close to each other, the quantum pressure can 
be larger than gravity to keep the accuracy of acceleration, velocity and 
position calculation. It turns out that all time steps are adaptively 
shortened than the one without 
quantum pressure.

\section{Softening length and kernel size}
The gravitational softening length (option \texttt{SofteningHalo} in \texttt{Gadget2}) 
is a critical numerical parameter for N-body simulation 
because it can avoid approaching the singularity 
when the distance between two particles becomes very small. 
In addition, some softening lengths within a certain volume can 
help to maintain the correct gravity force.
Ideally, if one could use infinite numbers of 
particles to represent the fluid,
then the softening length would be zero. 
However, in our FDM simulations, 
the softening length cannot be zero nor a completely artificial parameter because 
the density distribution is defined by the Gaussian kernel.
Unlike CDM, the FDM softening length is strongly related to the matter wavelength.  
In this appendix, we estimate the softening length as a
function of the wavelength $\lambda$.

\begin{figure}
\centering
\includegraphics[scale=0.5]{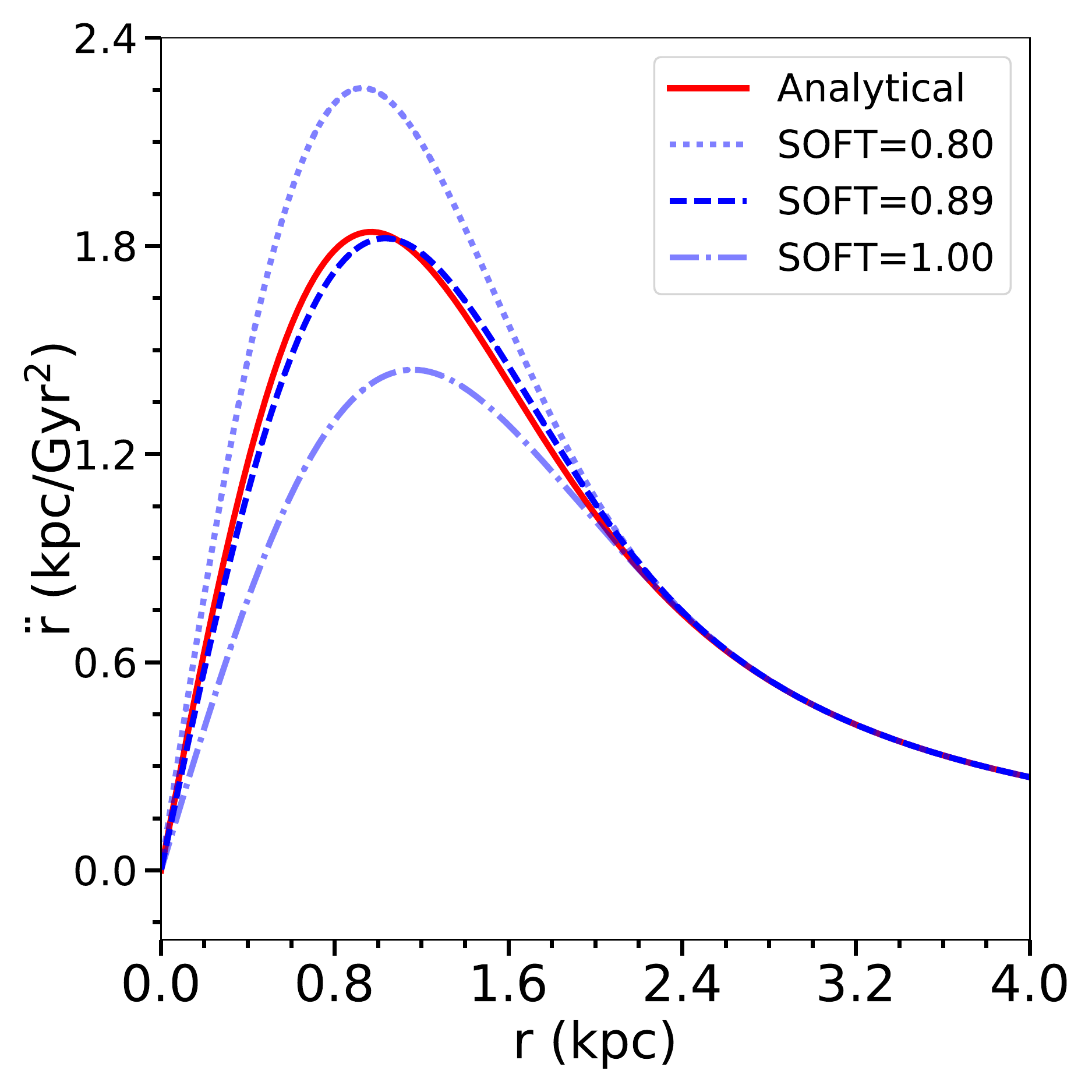}
\caption{The red curve is the analytical form of gravitational acceleration from a gaussian density distribution. 
The blue dashed curve is the best fit for $\texttt{SOFT}=0.89\kpc$, 
where \texttt{SOFT} is short for the softening length option \texttt{SofteningHalo}. 
The other two light blue curves are $\texttt{SOFT}=0.8\kpc$ (upper, dotted) and 
$\texttt{Soft}=1\kpc$ (lower, dot-dashed). 
}
\label{fig:kernel_gravity}
\end{figure}

For a distribution of mass with Gaussian density distribution, 
the gravitational acceleration acting on a test particle is 
\begin{equation}
\ddot{\boldsymbol{r}}=\dfrac{GM(<r)}{r^3}\boldsymbol{r},
\end{equation}  
where $M(<r)$ is the mass enclosed in the radius $r$ from the center of the Gaussian kernel.
The mass enclosed in the radius $r$ can be re-parameterized as   
\begin{equation}
\dfrac{M(<r)}{M(r=\infty)}=\dfrac{\int_{0}^{r}\exp(\dfrac{-2r^2}{\lambda^2})4\pi r^2 dr}{\int_{0}^{\infty}\exp(\dfrac{-2r^2}{\lambda^2})4\pi r^2dr}
={\rm erf}(\dfrac{\sqrt{2}r}{\lambda})-1.13\exp(\dfrac{-2r^2}{\lambda^2})\dfrac{\sqrt{2}r}{\lambda}.
\label{eq:Mlambda}
\end{equation}

In Fig.~\ref{fig:kernel_gravity}, we present various
gravitational acceleration 
(blue dashed lines) by taking into account different softening lengths, 
also see Eq. (4) in Ref.~\cite{gadget2}. 

We also show the analytical curve (red solid line) that 
is computed by assuming the
Gaussian kernel density distribution described in Eq.~\eqref{eq:Mlambda}. 
Clearly, the softening length $\texttt{SOFT}=0.89\kpc$ gives 
the best fit to the analytical curve, 
yet $\texttt{SOFT}=0.8\kpc$ or $\texttt{SOFT}=1\kpc$ 
differ by less than $50\%$. 
Therefore, we use the best-fit value of softening length $0.89\kpc$ in 
the simulation. 
The wavelength $\lambda=1.4$ kpc has been chosen in this work,
nevertheless, $\lambda$ and the softening length $\texttt{SOFT}=0.89\kpc$ 
are correlated in this context.

\section{Verification of Particle-Particle Simulation}
\label{sec:1D}

In this section, our PP method is verified with some simple one-dimensional simulations by comparing 
with other methods such as the PM method and the approach introduced in~\citep{veltmaat2016cosmological}. 
We first perform the simulations based on the PP and PM methods.
The same initial conditions are applied: two particles located 10 distance units away from each other 
and the density distribution smoothed by Gaussian kernels.  
For the sake of simplicity, we turn off the gravitational potential so that 
the acceleration of the particles is only caused by the QP.

In the PP simulation, we adopt the acceleration of particle described by Eq.~\eqref{eq:accel}  
and then evolve the particles with the iteration relations   
\begin{equation}\label{eq:evolve}
x_{t+\delta t} = x_t+v_t \delta t+\dfrac{1}{2}a_t (\delta t)^2,~~v_{t+\delta t}=v_t+\dfrac{1}{2}a_t \delta t,
\end{equation}
where $x_t$, $v_t$ and $a_t$ are the position, velocity and acceleration of the particle at time $t$, respectively. 
The size of the time step is given by an adjustable value of $\delta t$ 
in order to achieve adequate accuracy. 

In the PM simulation, the acceleration $a_t$ is $-\nabla Q$ based on the original definition.  
The acceleration $-\nabla Q$ shall be computed for each bin. 
To handle the $\nabla^2 \sqrt{\rho}$ in the QP given by Eq.~\eqref{pressure}, 
the three-point function is engaged to numerically calculate second order derivative,   
\begin{equation}
f''(x)=\lim_{h \to 0}\dfrac{f(x+h)-2f(x)+f(x-h)}{h^2}, \nonumber
\end{equation}
where $h$ is the bin size. 
However, we adopt the two-point function
\begin{equation}
f'(x)=\lim_{h \to 0}\dfrac{f(x+h)-f(x-h)}{2h}, \nonumber
\end{equation}
for calculating $-\nabla Q$. 
We have carefully checked the convergence of the simulations with different values of $\delta t$ and $h$. 
With the values $\delta t=0.01$ units of time and $h=0.05$ units of distance, 
the result is stable for illustration. 

\begin{figure}
\centering
\includegraphics[scale=0.5]{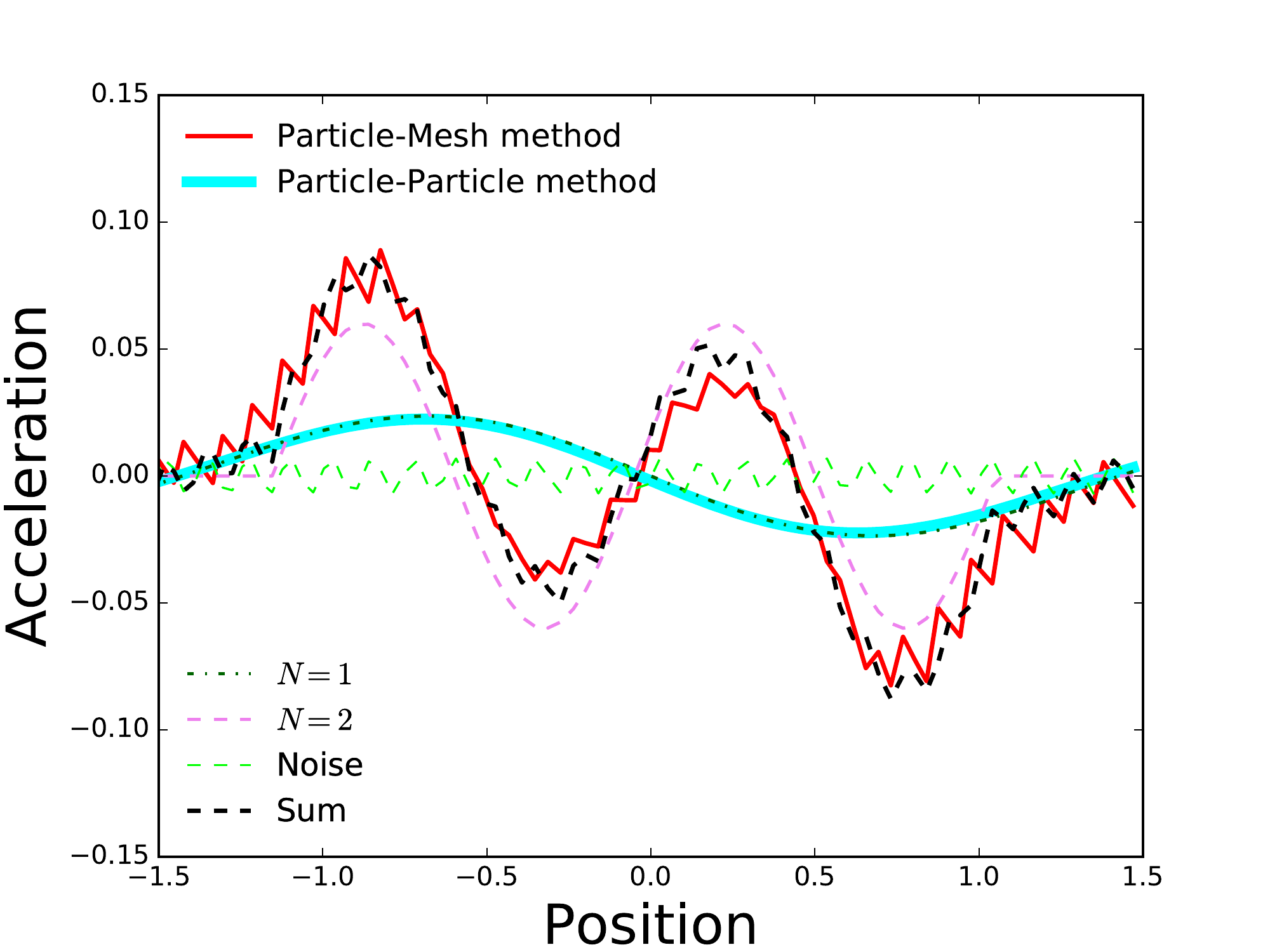}
\caption{Acceleration versus position of one particle in different situations. 
Here the thick red solid line is from the simulation with the PM method, 
while the thick cyan solid line is from the simulation with the PP method. 
The result of the simulation with the PM method can be decomposed into two modes ($N=1$ and $N=2$) together with numerical noise. 
The result of the simulation with the PP method is indeed the lowest frequency mode ($N=1$).
The dark green dot-dashed line, purple dashed line and light green dashed line are the lowest frequency mode ($N=1$), the high frequency mode ($N=2$) and the noise, respectively.
}
\label{fig:accelvspos_mode}
\end{figure}

The comparison of PP and PM results for the correlation between acceleration and position of a particle 
is demonstrated in Fig.~\ref{fig:accelvspos_mode}.
The thick red solid line presents the PM result and the thick cyan solid line is for the PP result.
One can decompose the PM simulation result into the lowest frequency mode ($N=1$), high frequency mode ($N=2$) 
and noise, corresponding to the dark green dot-dashed line, purple dashed line and light green dashed line, respectively.
From Appendix~B of Ref.~\citep{Hui:2016ltb}, 
we recognize that the lowest frequency mode is actually related 
to the lowest eigenstate of the Schr\"{o}dinger-Poisson equation (ground state $N=1$), and 
the contribution of the lowest eigenstate of Schr\"{o}dinger-Poisson equation is 
larger than other eigenstates in the density of the halo.  
Importantly, we found that the PP simulation result is dominated by the lowest frequency mode, 
which can be seen from the small difference between thick blue solid line and dark green dashed line. 

There is no doubt that the lowest frequency mode is more significant than 
higher frequency modes because the higher frequency modes are nearly averaged out in each interval.
This fact implies that our PP method for FDM simulation is only able to capture
the ground state component, nevertheless it is the most dominant component. 
Hence, such a PP method for FDM simulation is a good approximation and 
helps to reduce the simulation cost and complexity. 
There could be a very small error near the boundary of 
the softening length due to the omitted higher frequency modes when using our PP method. 
However, it is just a tiny effect on the large scale structure in a cosmological simulation 
because there is much weaker QP due to the larger simulation particle mass. 

\begin{figure}
\centering
\includegraphics[scale=0.5]{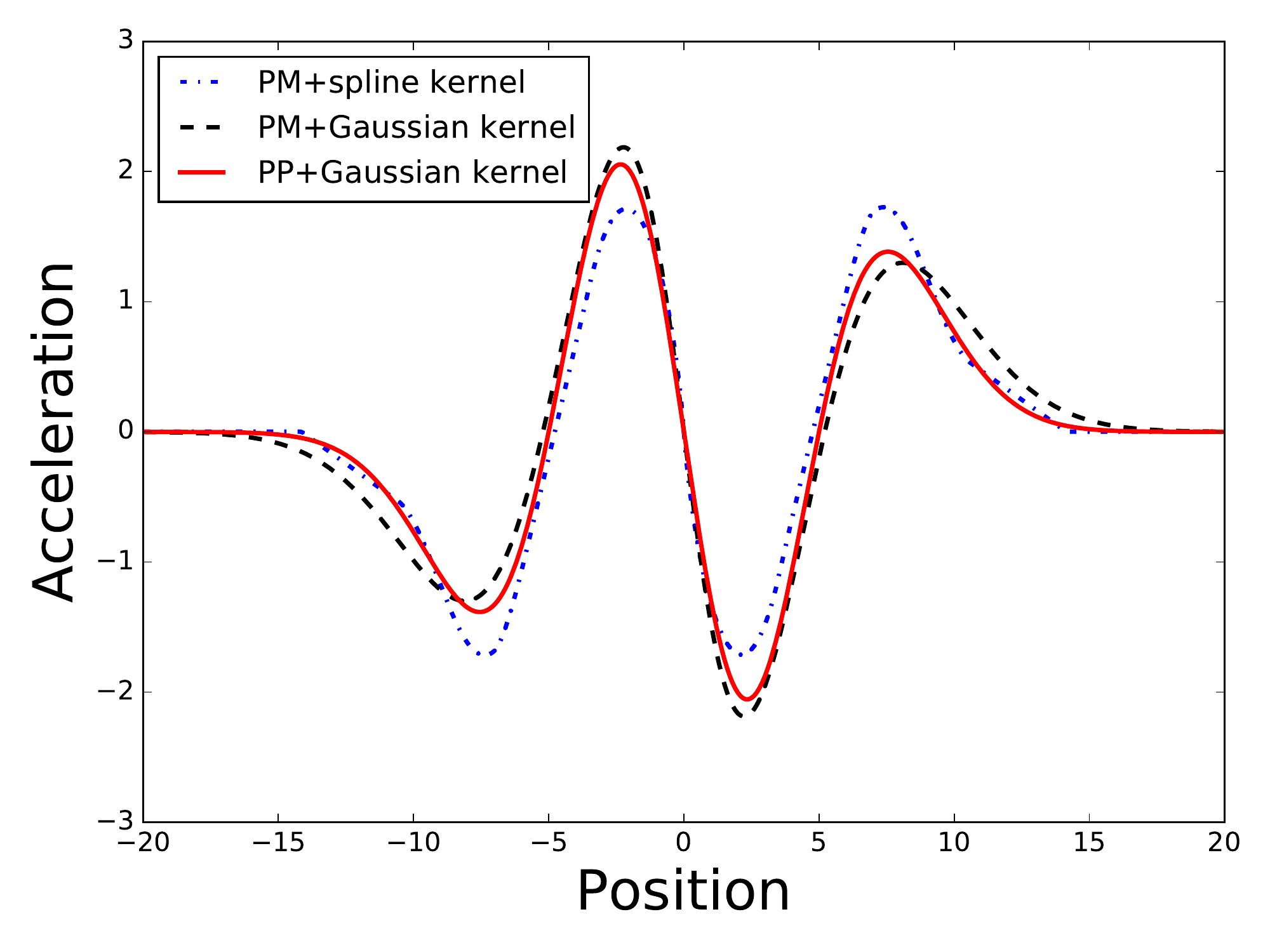}
\caption{Comparison of the result of acceleration from our method with the method in Ref.~\citep{veltmaat2016cosmological} and the usage of kernel in Ref.~\citep{veltmaat2016cosmological}.
The red solid line is the result of our PP method, the black dashed line is the result of the method in Ref.~\citep{veltmaat2016cosmological} with a Gaussian kernel and the blue dash-dotted line is the result of the method in Ref.~\citep{veltmaat2016cosmological} with a spline kernel (the kernel used in Ref.~\citep{veltmaat2016cosmological}).
}
\label{fig:accelvsposcomparethree}
\end{figure}

Next, we compare our PP method with the method developed in Ref.~\citep{veltmaat2016cosmological}.  
Again, we place two particles symmetrically and give them tiny initial velocity 
to make them move towards each other. 
We record the acceleration of one of the particles in the process 
as shown in Fig.~\ref{fig:accelvsposcomparethree}. 
In the simulation, we use the leapfrog algorithm to impose the acceleration 
on each particle and follow their movement.

The red solid line is the acceleration versus position curve recorded with the PP method.  
The black dashed and blue dash-dotted lines are recorded with the method developed in Ref.~\citep{veltmaat2016cosmological} 
but using the Gaussian and spline kernels, respectively.  
For convenience, we called the kernel used in Ref.~\citep{veltmaat2016cosmological} as 
spline kernel. 
Our PP method gives a similar acceleration versus position curve as that in Ref.~\citep{veltmaat2016cosmological}, 
despite of the different choices of the kernel. 
We also found that the Gaussian kernel is rather suitable for the QP calculation 
since the spline kernel is not smooth after performing first order derivative.

We conclude that our PP method is consistent with a previous PM method Ref.~\citep{veltmaat2016cosmological} 
in the aspect of QP calculation.
Moreover, our PP method captures the lowest frequency mode of the QP, whose resolution and accuracy are enough for a cosmological simulation. 
The further improvement of our method will lie in considering additional factors such as anisotropic kernel and higher frequency modes. 
We will study these possible improvements in the future.


\bibliography{MDM}


\end{document}